\documentstyle[prl,aps,twocolumn,graphicx,floats]{revtex}

\newcommand{\ud}{\,{\mathrm d}}

\begin{document}
\draft 
\title{Two-dimensional topological solitons in rectangular magnetic dots.} 
\author{Konstantin L. Metlov}
\address{Institute of Physics ASCR, Na Slovance 2, Prague 8, CZ-18221}
\date{\today} \maketitle
 
\begin{abstract}
  A general approach allowing to find the analytical expressions for
  equilibrium magnetic structures in small and flat magnetic
  nano-sized cylinders of arbitrary shape made of soft magnetic
  material is presented. The resulting magnetization distributions are
  two-dimensional topological solitons and have a non-zero topological
  charge. The approach is illustrated here on an example of a thin
  rectangular particle.
\end{abstract}
\pacs{75.75.+a, 75.60.Ch}

\paragraph*{Introduction.} The small magnetic cylindrical
elements (dots) made of a soft magnetic material became the focus of
attention due to their potential applications in the Magnetic Random
Access Memory (MRAM) devices. These elements can be made of different
shapes and sizes (usually of the order of $10-100 nm$ ) and found to
display a variety of magnetic structures (equilibrium distributions of
magnetization vector field). The structures are solutions of
non-linear and non-local (integral partial differential) equations of
micromagnetics and were a recent subject of extensive study by finite
element computer simulations (see e.g. \cite{HK99} and references
therein).

Recently, a new method allowing to treat the problem of finding the
magnetic structures of thin dots analytically became available
\cite{M01_solitons,M01_solitons2}.  It is based on the existence of
the well defined hierarchy of energies in small flat magnetic
cylinders made of soft magnetic material \cite{M01_solitons2}. The
resulting analytical magnetization distributions minimize the exchange
energy exactly, have no magnetic charges on the cylinder faces (except
in a few points of topological singularities) and no magnetic charges
on the cylinder sides.

Briefly, the general recipe of Ref. \cite{M01_solitons2} can be
formulated as follows. In a thin ferromagnetic cylinder (with thickness
$L \leq L_E$, where $L_E=\sqrt{C/M_S}$ is the exchange length, $C$ is
the exchange constant, $M_S$ is the saturation magnetization of the
material) the dependence of the magnetization distribution
$\vec{m}(\vec{r})=\vec{M}(\vec{r})/M_S$ on the coordinate $Z$
perpendicular to the cylinder face can be neglected (so that
$\vec{m}(X,Y)$, where $X$ and $Y$ are Cartesian coordinates on the
planes parallel to the cylinder face). Also, because the magnetization
vector has a fixed length $|\vec{m}|=1$ it has only two independent
components. Both these facts make it convenient to parametrize the
magnetization vector field by a complex function $w(z)$ of the complex
variable $z=X+\imath Y$, $\imath=\sqrt{-1}$, so that
$m_Z=(1-w(z)\overline{w(z)})/(1+w(z)\overline{w(z)})$, $m_X+\imath m_Y
= 2 w(z)/(1+w(z)\overline{w(z)})$, where a line over symbol means the
complex conjugation. The requirement of minimum face magnetic charges
makes the function $w(z)$ non-analytical (in the sense of the
Cauchy-Riemann conditions), but it is possible to express it
\cite{M01_solitons} through another, analytical, function $f(z)$. For
a particle with corners this can be done \cite{M01_solitons2} as
\begin{equation}
  \label{eq:sol_SM2}
  w(z)=\left\{
    \begin{array}{ll}
      f(z) & |f(z)| \leq 1 \\
      f(z)/\sqrt{f(z) \overline{f(z)}} & 1<|f(z)| \leq d\\
      f(z)/d & |f(z)| > d,
    \end{array}
    \right. 
\end{equation}
where $d$ is an arbitrary constant defining the size of regions of
the out-of-plane magnetization vector at topological singularities at
cylinder corners.

Let us denote the conformal transformation of the interior of the unit
circle $|t|\leq 1$ to $z \in \cal D$ ($\cal D$ is the set of points of
the $X-Y$ plane belonging to the cylinder face) as $z=T(t)$, according
to Riemann theorem such transformation exists for any simply connected
region $\cal D$. Then, the analytical function $f(z)$ corresponding
to the magnetization vector field with no magnetic charges on the
cylinder side is given \cite{M01_solitons2} by
\begin{equation}
  \label{eq:HP_sols}
  f(z) = (\imath t c + A - \overline{A} t^2) T'(t), \qquad 
  t\rightarrow T^{-1}(z),
\end{equation}
where $c$ and $A$ are arbitrary real and complex constants
respectively, $T^{-1}(z)$ is the inverse function of $T(t)$. The
function $T^{-1}(z)$ exists and is unique because the transformation
is conformal. The expression (\ref{eq:HP_sols}) contains essentilly
all the solutions minimizing the exchange energy and having no
magnetic charge on the cylinder sides. The magnetostatic energy of the
resulting solutions is also very small compared to the exchange energy
(but this depends on the particle size) and it can be expected (for
sufficiently small particles) not to introduce any new solution types
but just to modify (\ref{eq:HP_sols}) slightly.

\begin{figure}[htbp]
  \begin{center}
    \includegraphics[scale=0.65]{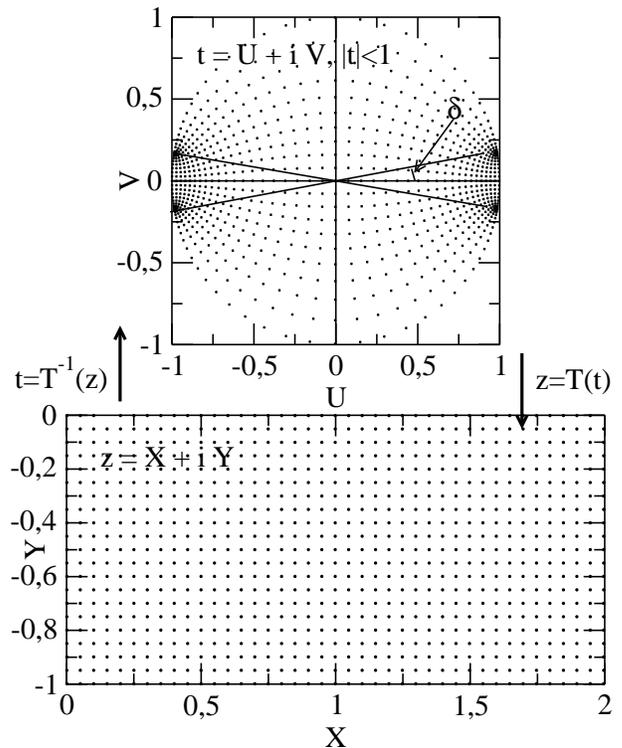}
    \caption{Conformal mapping of the unit circle to the rectangle, 
      dots on both plots correspond to each other to show how the
      interior is transformed.}
    \label{fig:conform}
  \end{center}
\end{figure}

\paragraph{The magnetic structures in a rectangular particle.} The shape
of the particle enters the resulting distribution through the
conformal transformation $T(t)$. The conformal transformation of the
unit circle to the rectangle is
\begin{equation}
  \label{eq:Tt}
  T(t)= C_1 (\int_0^t \frac{1}{\sqrt{1+t^4-2 t^2 \cos 2 \delta}} 
  \ud t + C_0),
\end{equation}
which is a particular case of Schwartz-Christoffel integral. This
integral can be expressed analytically through the elliptic integral
of the first kind $F(\phi,m)$. The parameter $0\leq\delta\leq\pi/4$
controls the aspect ratio of the rectangle, $\delta=\pi/4$ corresponds
to square, $\delta \simeq 0.172426$ to the $2:1$ rectangle. The
constants $C_1$ and $C_0$ were chosen in such a way that the height of
the rectangle is $1$ and its upper left corner is situated at the
point with $X,Y=0$. For the $2:1$ rectangle shown in the pictures $C_1
\simeq 0.63189$ and $C_0\simeq -1.849 + \imath 0.787$. This
transformation is visualized in Fig.~\ref{fig:conform}.

The magnetic structures in the particle are thus
\begin{equation}
  \label{eq:HP_sols_rect}
  f(z) = (\imath t c + A - \overline{A} t^2) 
  \frac{1}{\sqrt{1+t^4-2 t^2 \cos 2 \delta}}, \; t\rightarrow T^{-1}(z),
\end{equation}
where $C_1$ is absorbed into $c$ and $A$.

To find the values of constants $c$ and $A$ it is required to minimize
the total energy of the particle (including magnetostatics). This will
be done elsewhere, let us now just treat these constants as
independent parameters and analyze the possible types of solutions
qualitatively.

\begin{figure}[htbp]
  \begin{center}
    \includegraphics[scale=0.48]{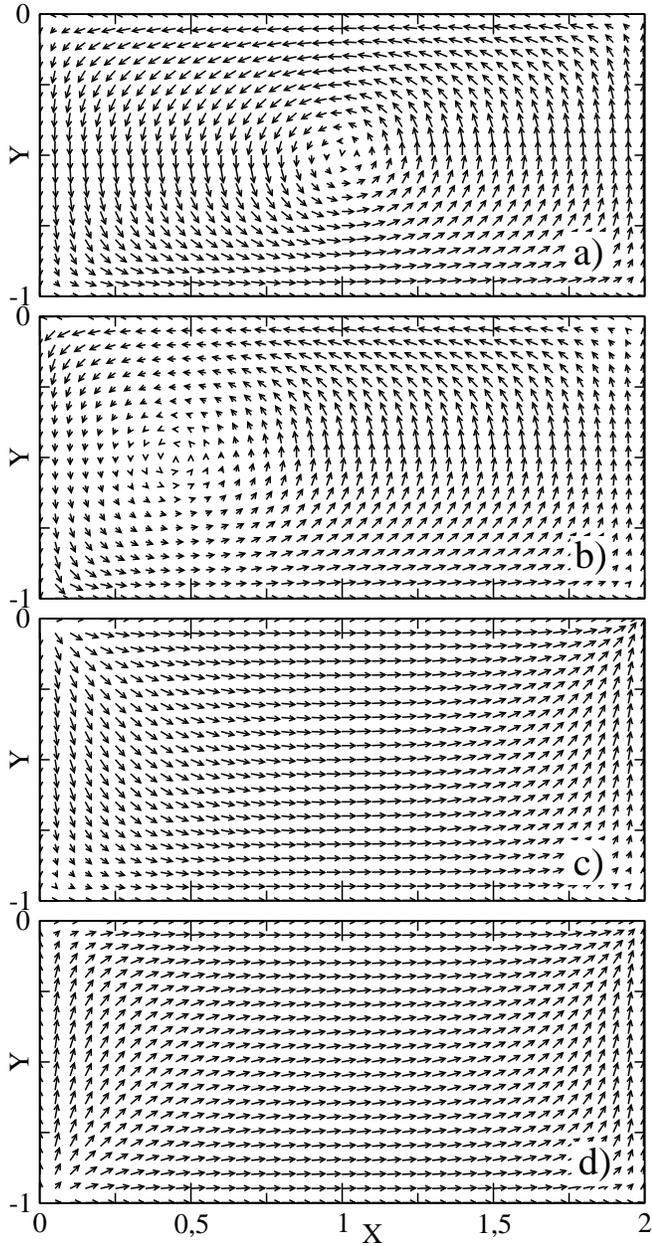}
    \caption{The equilibrium configurations of the magnetization
      vector (its projection to the cylinder face is shown) in the
      rectangular particle (\ref{eq:sol_SM2}), (\ref{eq:HP_sols_rect})
      with $d=6$: a) vortex $c=2$, $A=0$, ; b) displaced vortex $c=2$,
      $A=1.9\imath$; c) ``C'' structure $c=2$, $A=11.6$; d) ``S''
      structure $c=0$, $A=6 e^{\imath 0.172426}$.}
    \label{fig:structures}
  \end{center}
\end{figure}

As it was already shown \cite{M01_solitons2} for the cases of circular
and triangular cylinders (and it is applicable to any soliton of type
(\ref{eq:HP_sols})) there are two basic classes of solutions depending
on whether $|c|<2|A|$ or not. The first class, corresponds to the case
$|c|>2|A|$ is a vortex, the sign of $c$ controls the direction of the
rotation of the magnetization vector in it. If the particle is square
the centered vortex (Fig~\ref{fig:structures}a) is stable. In
elongated particles a state with displaced vortex
(Fig~\ref{fig:structures}b, see also Figs~3,5 in Ref.~\cite{HK99}) can
also be. The other class of solutions contained in (\ref{eq:HP_sols})
corresponds to the two skyrmions (hedgehogs) bound to the cylinder
sides. If the particle is not of the circular shape positions of the
hedgehogs on its boundary are not equivalent.  Particularly, for
particle with corners the total energy is lower if the hedgehogs are
located in the corners (however states with hedgehogs at the middle of
the edges can be metastable). For rectangle (because of the symmetry)
there are two possibilities: either the hedgehogs are located at the
neighbouring corners (Fig~\ref{fig:structures}c), or at the opposite
ones (Fig~\ref{fig:structures}d). Both these structures, usually
called ``C'' and ``S'', already appeared in numerical micromagnetic
simulations for rectangles.

Because of the completeness of the solution (\ref{eq:HP_sols}) no
other shapes of the equilibrium magnetization vector field can be
expected in the rectangular particle if it is sufficiently small (but
larger then the single domain size, where absence of magnetic charges
is not required) and if it is made of sufficiently isotropic magnetic
material, which are the basic assumptions of the presented theory.

The expression (\ref{eq:HP_sols_rect}) also describes the metastable
states when the $c$ and $A$ deviate from their equilibrium values.
These states can be useful for considering the magnetization reversal,
excitations and the stability of the equilibrium solutions (see e.g.
\cite{GM01}).

\paragraph*{Summary.} A simple analytical framework for finding the
profiles of two-dimensional topological solitons in small flat
ferromagnetic cylinders of the arbitrary shape was presented. As an
example, the case of the rectangular cylinder was considered.

The resulting magnetization vector fields may find their use in
magnetic microscopy (for fitting the signal from the probe by changing
the parameters $A$ and $c$), and for building the phase diagrams of
small magnetic particles outlining the regions of the particle
geometrical parameters where each of the presented equilibrium
solutions correspond to the ground state of the system.

This work was supported in part by the Grant Agency of the Czech
Republic under projects 202/99/P052 and 101/99/1662. I would like to
thank Ivan Tom{\'a}{\v s} for reading the manustcript and many
valuable discussions.

\end{document}